\documentclass[superscriptaddress,onecolumn,pra,preprintnumbers,amsmath,amssymb,showpacs]{revtex4}
\usepackage{lipsum} 
\usepackage[latin9]{inputenc}
\usepackage{graphicx}
\usepackage{dcolumn}
\usepackage{bbold}
\usepackage{bm}
\usepackage[usenames]{xcolor}

\usepackage[colorlinks,bookmarks=false,citecolor=blue,linkcolor=blue,urlcolor=blue,hyperfootnotes=true]{hyperref}

\makeatletter
\newsavebox\myboxA
\newsavebox\myboxB
\newlength\mylenA

\newcommand*\xoverline[2][0.75]{%
    \sbox{\myboxA}{$\m@th#2$}%
    \setbox\myboxB\null
    \ht\myboxB=\ht\myboxA%
    \dp\myboxB=\dp\myboxA%
    \wd\myboxB=#1\wd\myboxA
    \sbox\myboxB{$\m@th\overline{\copy\myboxB}$}
    \setlength\mylenA{\the\wd\myboxA}
    \addtolength\mylenA{-\the\wd\myboxB}%
    \ifdim\wd\myboxB<\wd\myboxA%
       \rlap{\hskip 0.5\mylenA\usebox\myboxB}{\usebox\myboxA}%
    \else
        \hskip -0.5\mylenA\rlap{\usebox\myboxA}{\hskip 0.5\mylenA\usebox\myboxB}%
    \fi}
\makeatother

\begin{document}

\title{Orbital selectivity versus Pomeranchuk instability in the iron-chalcogenide superconductors: A two-loop renormalization group study}

\author{Rafael R. Caetano}
\affiliation{Instituto de Física, Universidade Federal de Goiás, 74.001-970, Goiânia-GO, Brazil}
\author{Hermann Freire$^1$}

\date{\today}
\begin{abstract}
We perform a two-loop renormalization group (RG) analysis of a 2D effective multiband model, which is relevant for
describing the low-energy properties of some iron-chalcogenide superconducting materials. Crucial ingredients in this analysis are the calculation of higher-order contributions in the RG scheme
that go beyond the widely-used parquet approximation and the consequent inclusion of nontrivial self-energy effects of the model
that yield an anisotropic renormalization of the quasiparticle weight in the system. The motivation of our work is the experimental discovery by Sprau \emph{et al.} [Science \textbf{357}, 75 (2017)]
that orbitally-selective renormalization of the quasiparticle weight in the Hund's metal phase at moderate temperatures underpins the highly unusual gap in the superconducting phase of the FeSe compound at lower temperatures.
One prediction we arrive here is that the underlying origin of nematicity in these systems may indeed come from orbital-selectivity, instead of a Pomeranchuk instability in the $d_{\pm}$ channel. This orbital selectivity is driven by the presence of stripe-type antiferromagnetic fluctuations in the model. Therefore, we argue that the present RG results may provide a scenario from a weak-to-moderate coupling perspective, in which the role of orbital selectivity to describe the physical properties of some iron-chalcogenide superconductors is emphasized. 
\end{abstract}


\maketitle

\section{Introduction} 

The highly unconventional properties recently observed in the iron-based chalcogenides superconductors are currently at the center stage in the field of strongly correlated systems \cite{Sprau,Hirschfeld,Kreisel,Kostin}.
One of the primary reasons for this is due to the fact that the underlying mechanism of superconductivity, which manifests itself in these compounds, fails to follow the ``typical'' behavior commonly observed in several other iron-based superconductors \cite{Hosono,Greene,Jonhston,Stewart}.
Indeed, while in most of these compounds superconductivity always emerges close to both an antiferromagnetic order and a nematic phase (the latter being associated with a spontaneous breaking of the $C_4$ rotational symmetry down to $C_2$), only a nematic phase is observed in the FeSe compound with no magnetic order at ambient pressure \cite{Cava}. This has led some researchers to question on a microscopic level if the formation of Cooper pairs in these latter materials is only spin-fluctuation mediated or is a result of a much more complex interplay of magnetic, structural and orbital degrees of freedom.

The nematic transition in bulk FeSe has a critical temperature \cite{Cava}  of $T_s=90$ K and the corresponding superconducting critical temperature \cite{Wu} around $T_c=8$ K, which seems at first very low. However,
thin films of this class of materials offer a unique opportunity to easily change $T_c$ and increase its value by at least one order of magnitude such as, e.g., in single layers of FeSe grown on SrTiO$_3$ \cite{Sadovskii,Wang,Hoffman,Jia}.
This evidence probably indicates an unconventional mechanism at work in these materials. This discovery may open a new avenue to engineer new superconducting materials with even higher critical temperatures \cite{Shi}, which can be of course very interesting for technological applications. 

From a fundamental point of view, a recent groundbreaking work has shed new light on this problem. Indeed, state-of-the-art scanning tunneling microscopy (STM) experiments performed by the group of Séamus Davis with collaborators \cite{Sprau,Hirschfeld} have provided solid evidence of an orbitally-selective superconductivity which emerges from a novel Hund's metal phase with anisotropically renormalized quasiparticle weights at the different orbitals that make up the underlying Fermi pockets of these systems. This remarkable experimental discovery indicates a profound role of the strong correlations present in these systems that must be included in a microscopic theory that seeks to describe the mechanism of 
the unconventional superconductivity that manifests itself in this family of materials.

The complex interconnection of many different phases (i.e., orbital order, antiferromagnetism and superconductivity) in these materials calls for a theoretical approach which describes these highly entangled 
quantum fluctuations that appear in these systems on equal footing. Renormalization group (RG) methods in general are an ideal approach to
attack these types of problems (see, e.g., Refs. \cite{Freire,Freire2,Shankar}), in view of its completely unbiased nature. Recently, one important RG result in the literature was obtained by Chubukov \emph{et al.} \cite{Chubukov} who analyzed an itinerant multiband model with a three-orbital electronic structure with both Hubbard and Hund's interactions relevant for describing the iron-based superconductors \cite{Chubukov2} within a one-loop parquet RG scheme. Remarkably, these authors have
obtained that there is indeed a regime in this model in which spontaneous orbital order becomes the dominant instability at very low temperatures
as compared to a $s_{\pm}$ singlet pairing state (this latter order implies a sign-change in the superconducting gap between electron and hole pockets). In addition, this orbital order also acts to prevent the system from ordering antiferromagnetically. 
As a consequence, they conjectured that this scenario may be relevant as a possible explanation as to why no magnetic order is observed in the FeSe compound at ambient pressure. 
Complementary numerical results on a slightly different multiband model with the addition of more ingredients were independently obtained by Honerkamp and confirmed the emergence of a spontaneous orbital order for large interorbital interactions in the iron-chalcogenide systems using one-loop functional RG methods \cite{Honerkamp} .

A key aspect of this problem we would like to emphasize here is that, broadly speaking, there are two point of views that compete for the qualitative understanding of the FeSe compound. These theories can be roughly divided into weak-coupling and strong-coupling approaches. In the former case,
we point out that recently a phenomenological spin-fluctuation pairing theory has been proposed in the literature \cite{Kreisel} that incorporates the lack of coherence of some low-energy quasiparticles at different orbitals (thus, emphasizing the concept of orbital selectivity), which successfully describes the highly anisotropic gap structures that are experimentally measured inside the superconducting phase of these compounds \cite{Sprau}. On the other hand, it is also important to mention that it has been argued recently that, due to the smallness of the Fermi pockets displayed by the FeSe compound, this system may be possibly in a more strongly coupled regime, within the realm of a multiband BCS-BEC crossover \cite{Matsuda,Matsuda2,Chubukov3}. In this respect, other approaches which emphasize the importance of the strong interactions present in these systems have also appeared in the literature. For instance, in Ref. \cite{Si}, a strongly coupled model with localized spins was put forward and has been also applied to describe the FeSe compound. 

\begin{figure}
\includegraphics[width=1.8in]{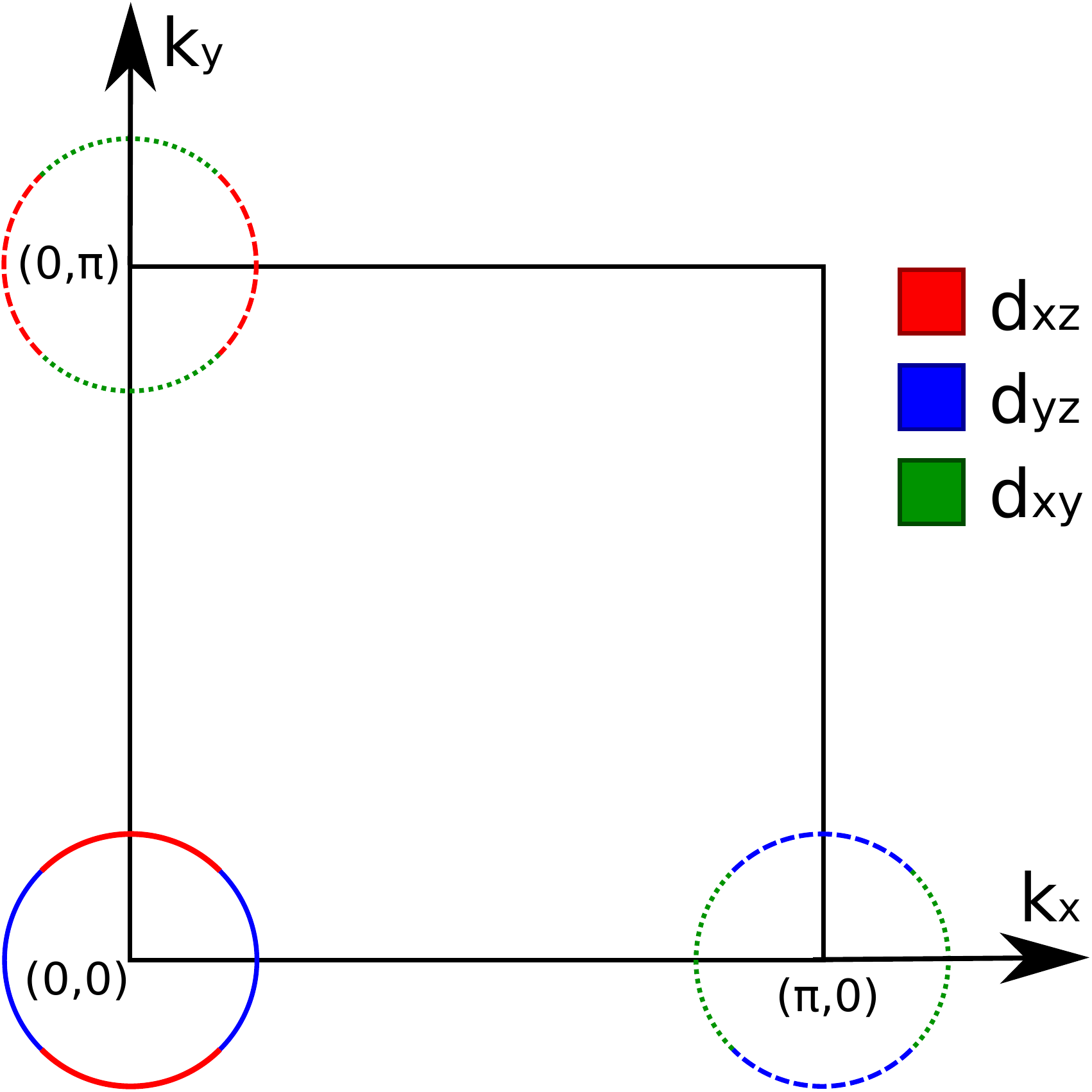}
\caption{(Color online) The Fermi pockets in the present 2D multiband model: A small hole pocket around $\Gamma=(0,0)$ and small electron pockets around $X=(\pi,0)$ and $Y=(0,\pi)$, which also includes the orbital components ($d_{xy}$, $d_{xz}$, and $d_{yz}$). The hole pocket is approximately nested with the electron pockets, where the nesting wavevectors are given by $\mathbf{Q_{x}}=(\pi,0)$ and $\mathbf{Q_{y}}=(0,\pi)$. Since our primary aim in this work is to discuss the $xz/yz$ splitting, the orbital $d_{xy}$ is neglected for simplicity.}
\end{figure}

Moreover, from a broad perspective, Hund's metals \cite{Kotliar} emerge as novel unconventional electronic phases, in which the underlying quasiparticles display a dichotomic character: They exhibit both itinerant and localized natures that coexist at the Fermi pockets of these systems \cite{Kotliar}. For this reason, this new concept attracted
considerable interest in the field of superconductivity in the iron-based compounds since it implies a new metallic state, which is halfway between a conventional metal and a Mott insulator (for an excellent review on Hund's metal, see, e.g., Ref. \cite{Georges}). Therefore, the electron correlations in this phase are naturally expected to be of intermediate strength and, in view of the mutiband nature of the problem, turn out to be not strong enough to drive the system into insulation (as it normally occurs in single-band models). Another interesting aspect of this unusual metallic phase is that it is often unstable at lower temperatures towards becoming unconventional superconductors with enhanced critical temperatures. This fact highlights the importance of Hund metallicity, as it implies including new ingredients in order to explain, even at a qualitative level, the underlying mechanism of the Cooper-pair formation in many iron-based systems.

\begin{figure*}
\includegraphics[width=4.5in]{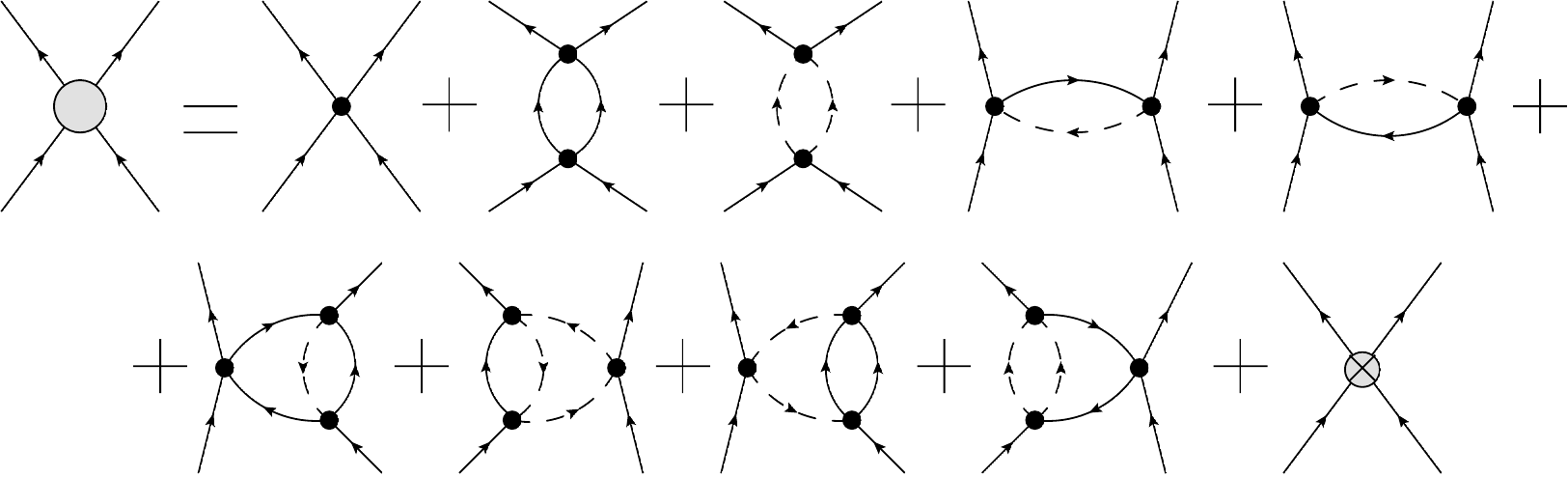}
\caption{Schematic representation of the Feynman diagrams in the RG calculation of the vertex corrections up to two-loop order, e.g., in one interaction channel (for the other channels, the derivation of the corresponding diagrams is straightforward). The solid lines refer to the noninteracting Green's function associated with excitations in the vicinity of the hole pocket, whereas the dashed lines stand for the noninteracting Green's function associated with excitations close to the electron pockets in the present 2D multiband model. The last diagram with a cross represents the corresponding counterterm, which regularizes order by order the renormalized perturbation theory.}
\end{figure*}

In this work, we implement a two-loop RG approach for a minimal multiband model, which includes a multiorbital structure in order to describe qualitatively the FeSe compound. From a technical standpoint, the two-loop RG calculation in this model turns out to be necessary in view of the fact that, at this order, it is possible to include nontrivial self-energy effects consistently into the RG scheme. For this reason,
we will be able to investigate the strength of the correlations by carefully monitoring the renormalization of the quasiparticle weight in a orbitally-selective way in the present model. Interestingly, we will show that there is indeed an unconventional metallic regime, in which the quasiparticle excitations with different degrees of low-energy coherence associated with different orbital components coexist, which will have an impact on the symmetry-broken phases that potentially emerge at lower temperatures in the model. We then proceed to analyze the physical consequences and the limitations of this RG theory for describing the FeSe compound.

This paper is structured as follows: First, we define a minimal multiband model, which includes initially only the orbitals $d_{xz}$ and $d_{yz}$ as a possible starting point seeking to provide a qualitative description of some iron-based superconductors at moderate temperatures. Then, we proceed to perform a one-loop RG calculation that describes the flow of the coupling constants of the model as one integrates out progressively the high-energy
degrees of freedom towards its low-temperature limit in order to emphasize the importance of correlations in this problem. Next, we move on to the two-loop RG calculation that includes both vertex corrections and nontrivial self-energy effects to determine consistently how the quasiparticle weights at different orbitals renormalize within a weak-to-moderate coupling scenario. Finally, we conclude by comparing all the results obtained here both with experimental observation and with other theoretical works available in the literature.

\section{Minimal multiband model}

As explained before, we will focus here specifically on the case of the FeSe compound. The Fermi surface shown by ARPES experiments \cite{Suzuki,Coldea,Zhou,Kordyuk}, quantum oscillations \cite{Terashima,Coldea2} and scanning tunneling spectroscopy \cite{Kostin,Kreisel,Hanaguri} of the metallic phase of this system in the one-iron unit cell zone comprises of many pockets that display a rich multiorbital content: two small hole pockets around $\Gamma=(0,0)$ associated with $d_{xz}$ and $d_{yz}$ orbitals, one small electron pocket around $X=(\pi,0)$ containing $d_{xy}$ and $d_{yz}$ orbitals, and one small electron pocket around $Y=(0,\pi)$ associated with $d_{xy}$ and $d_{xz}$ orbitals (see Fig. 1). 
The degeneracy of the two hole pockets around the $\Gamma$-point turn out to be split due to spin-orbit coupling \cite{Borisenko,Watson}. In this way, we take into account only one hole pocket in a minimal model to describe the FeSe compound.
We will depart here from
an itinerant 2D electronic model with three pockets as shown schematically in Fig. 1. Since our goal here is to perform a two-loop RG calculation, we will neglect, for simplicity, the 
$d_{xy}$ orbital \cite{Chubukov4}, in view of the fact that recent experiments have suggested that the quasiparticle weight associated with this latter orbital
is strongly suppressed relative to the $d_{xz}$ and $d_{yz}$ orbitals (\emph{i.e.}, the $d_{xy}$ is almost localized, while the $d_{xz}$ and $d_{yz}$ orbitals are more itinerant). We point out that we plan to include in a future work also the contribution of the $d_{xy}$ orbital within the present framework. Therefore, the noninteracting Hamiltonian associated with the hole pocket in the band basis is given by

\vspace{-0.2cm}

\begin{align}
&\mathcal{H}^{\Gamma}_{0} =\sum_{\mathbf{k}} \varepsilon_{c}(\mathbf{k}) d_{1,\mathbf{k}}^{\dagger}d_{1,\mathbf{k}},\label{Eq_01}
\end{align}

\noindent where $\varepsilon_{c}(\mathbf{k})=\mu-\mathbf{k}^2/(2m_c)$ with $\mu$
being the chemical potential and the wavevector $\mathbf{k}$ is near the $\Gamma$-point in the Brillouin zone. The noninteracting Hamiltonian associated with the electron pockets in the band basis becomes

\vspace{-0.2cm}

\begin{align}
&\mathcal{H}^{X,Y}_{0} =\sum_{\mathbf{k}} [\varepsilon_{f_1}(\mathbf{k}) f_{1,\mathbf{k}}^{\dagger}f_{1,\mathbf{k}}+\varepsilon_{f_2}(\mathbf{k}) f_{2,\mathbf{k}}^{\dagger}f_{2,\mathbf{k}}],\label{Eq_02}
\end{align}

\noindent where $\varepsilon_{f_1}(\mathbf{k})=\varepsilon_0+k_{x}^2/(2m_x)+k_{y}^2/(2m_y)-\mu$ is the dispersion
for excitations near the $X$-point in the Brillouin zone and $\varepsilon_{f_2}(\mathbf{k})=\varepsilon'_0+k_{x}^2/(2m_y)+k_{y}^2/(2m_x)-\mu$
is the dispersion for excitations close to the $Y$-point. We assume here that the hole and electron pockets are approximately nested with the nesting wavevectors given by $\mathbf{Q_{x}}=(\pi,0)$ and $\mathbf{Q_{y}}=(0,\pi)$ and $\mathbf{k}$ in the dispersions $\varepsilon_{f_1}(\mathbf{k})$ and $\varepsilon_{f_2}(\mathbf{k})$ is measured, respectively, with respect to $\mathbf{Q_{x}}$ and $\mathbf{Q_{y}}$, with the parameters $\varepsilon_0$ and $\varepsilon'_0$ being the corresponding energy shifts in the dispersions. Since our RG results will not depend qualitatively on the particular choice of the masses $m_c$, $m_x$ and $m_y$, we will set them henceforth equal to each other, i.e., $m_c=m_x=m_y=m$. Here, we also assume for simplicity that the bands of the system do not possess a van Hove singularity in their band structure. In addition, we will consider an effective multiband model that assumes as a starting point a constant density of states at the Fermi level of the system.

For the interacting part of the Hamiltonian, we will analyze here a model with both Hubbard and Hund's interactions, which is given in the orbital basis by

\vspace{-0.2cm}

\begin{align}\label{Eq_03}
\mathcal{H}_{int} &=U_1\sum_{\sigma\sigma'} f_{1\sigma}^{\dagger}d_{1\sigma'}^{\dagger}d_{1\sigma'}f_{1\sigma}+\bar{U}_1\sum_{\sigma\sigma'} f_{2\sigma}^{\dagger}d_{1\sigma'}^{\dagger}d_{1\sigma'}f_{2\sigma}\nonumber\\
&+U_2\sum_{\sigma\sigma'} f_{1\sigma}^{\dagger}d_{1\sigma'}^{\dagger}f_{1\sigma'}d_{1\sigma}+\bar{U}_2\sum_{\sigma\sigma'} f_{2\sigma}^{\dagger}d_{1\sigma'}^{\dagger}f_{2\sigma'}d_{1\sigma}\nonumber\\
&+\frac{U_3}{2}\sum_{\sigma\sigma'} [f_{1\sigma}^{\dagger}f_{1\sigma'}^{\dagger}d_{1\sigma'}d_{1\sigma}+H.c.]\nonumber\\
&+\frac{\bar{U}_3}{2}\sum_{\sigma\sigma'} [f_{2\sigma}^{\dagger}f_{2\sigma'}^{\dagger}d_{1\sigma'}d_{1\sigma}+H.c.]\nonumber\\
&+U_4\sum_{\sigma\sigma'} d_{1\sigma}^{\dagger}d_{1\sigma'}^{\dagger}d_{1\sigma'}d_{1\sigma}\nonumber\\
&+\frac{U^{(1)}_5}{2}\sum_{\sigma\sigma'} f_{1\sigma}^{\dagger}f_{1\sigma'}^{\dagger}f_{1\sigma'}f_{1\sigma}+\frac{U^{(2)}_5}{2}\sum_{\sigma\sigma'} f_{2\sigma}^{\dagger}f_{2\sigma'}^{\dagger}f_{2\sigma'}f_{2\sigma}\nonumber\\
&+\frac{\bar{U}_5}{2}\sum_{\sigma\sigma'} [f_{1\sigma}^{\dagger}f_{1\sigma'}^{\dagger}f_{2\sigma'}f_{2\sigma}+(1\leftrightarrow 2)]\nonumber\\
&+\tilde{U}_5\sum_{\sigma\sigma'} f_{1\sigma}^{\dagger}f_{2\sigma'}^{\dagger}f_{2\sigma'}f_{1\sigma}+\tilde{\tilde{U}}_5\sum_{\sigma\sigma'} f_{1\sigma}^{\dagger}f_{2\sigma'}^{\dagger}f_{1\sigma'}f_{2\sigma}\nonumber,\\
\end{align}

\noindent where it is assumed that momentum conservation holds in the above interactions (modulo a reciprocal wavevector in the case of Umklapp couplings) and the volume $V$ of the system has been set to unity. 
The above couplings are the bare ones, i.e. they are defined at a microscopic scale. For this reason, they must
be initially given by $U_{1,0}=U_{2,0}=U_{3,0}=U_{4,0}=U^{(1)}_{5,0}=U^{(2)}_{5,0}=U$ (where $U$ is the local Hubbard interaction between fermions at the same orbital),
$\bar{U}_{1,0}=\tilde{U}_{5,0}=U'$ (where $U'$ is the local Hubbard interaction between fermions at different orbitals),
$\bar{U}_{2,0}=\tilde{\tilde{U}}_{5,0}=J$ (where $J$ is the Hund's interaction between fermions at the same orbital),
and $\bar{U}_{3,0}={\bar{U}}_{5,0}=J'$ (where $J'$ is the Hund's interaction between fermions at different orbitals).
We point out that a similar itinerant 2D model was recently studied within a one-loop RG approximation in Ref. \cite{Chubukov}, in which these authors consider a low-energy effective four-pocket model to describe on a qualitative level many iron-based compounds.
Therefore, in order to allow an easy comparison between all the results available in the literature, we will follow here a similar RG strategy and the same notation as in Ref. \cite{Chubukov}. Indeed, as explained thoroughly in the Ref. \cite{Chubukov}, there is a difference of our couplings in the band basis compared to the orbital basis. While the orbital content of the excitations does not appear in the non-interacting part of the Hamiltonian, it does introduce angular dependencies in the four-fermion interaction terms in the band basis. Our defined 12 interaction parameters in the orbital basis turn out to be the prefactors of combinations, which result in many more couplings in the band basis. This makes the underlying model in the band basis completely $C_4$-symmetric, even though in the orbital basis the theory departs from the couplings $U$, $U'$, $J$ and $J'$. Interestingly, when we write down the flow equations for the couplings and the quasiparticle weight in the approximation $m_c=m_x=m_y=m$, the angular dependencies disappear altogether from the RG equations. In this way, we may directly analyze those flow equations in the orbital basis, which is what we do in the present work.

\begin{figure}[t]
\includegraphics[width=3.15in]{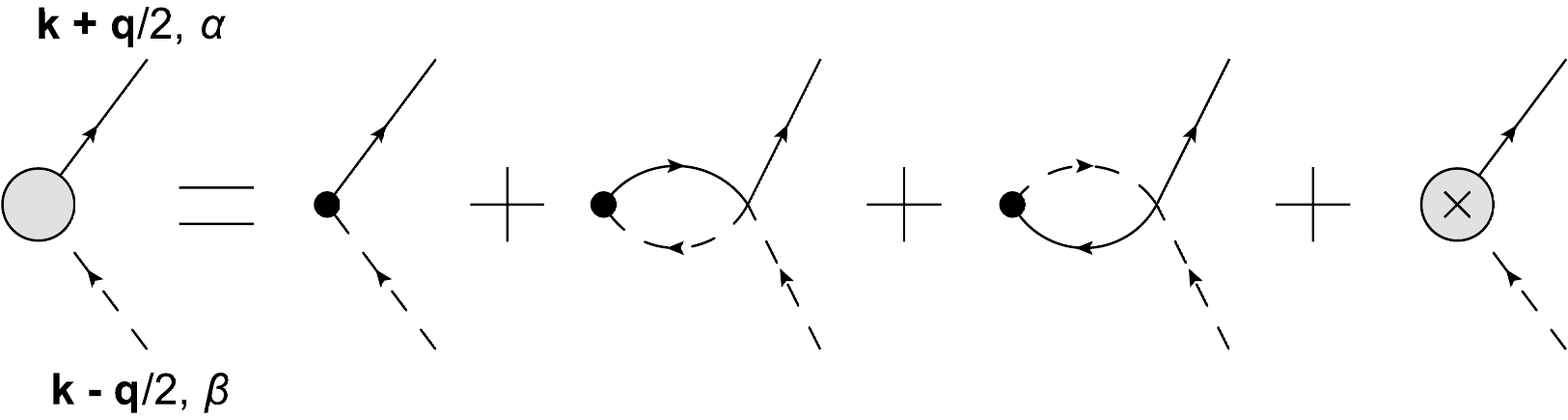}\\
\includegraphics[width=3.15in]{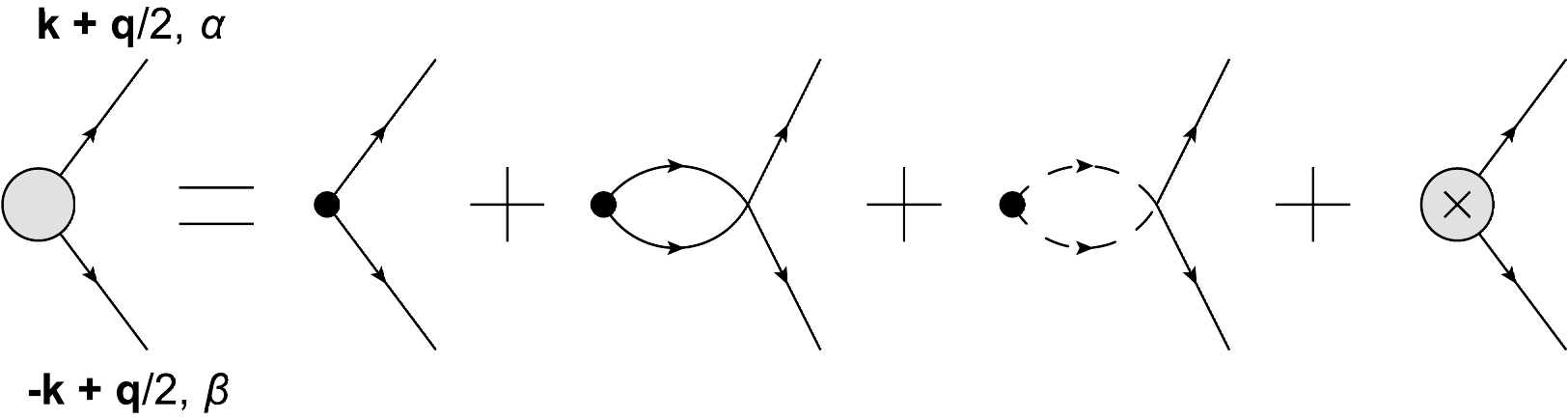}\\
\includegraphics[width=3.15in]{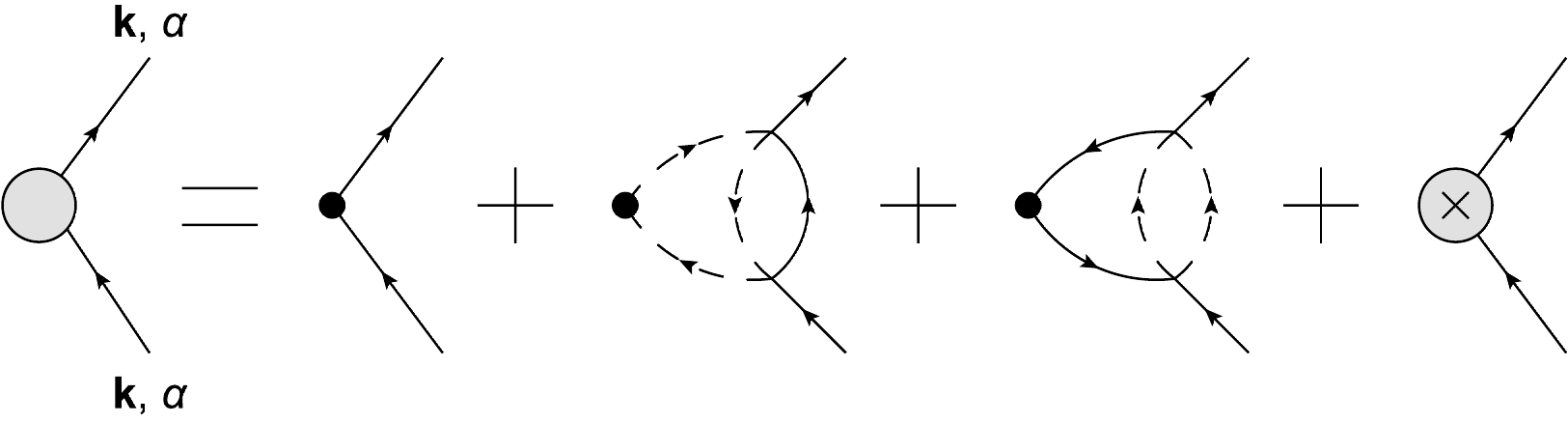}
\caption{(Color online) Feynman diagrams associated with the response vertices calculated in the present multiband model. The diagrams with crosses denote the corresponding counterterms that are defined in the text, which regularize the renormalized perturbation theory. The two-loop diagrams represent the so-called Aslamazov-Larkin diagrams.}
\end{figure}

The perturbative RG approach that we will adopt here is standard \cite{Peskin} and was explained in detail by one of the authors in the context of several different electronic models elsewhere \cite{Freire, Freire2,Freire3,Freire4}. If we apply a naive perturbation theory to the model defined in Eqs. (\ref{Eq_01})-(\ref{Eq_03}), 
divergences or non-analyticities will appear in the low-energy limit. We get around this problem by redefining all the couplings in the following way: $u_{i,0}=\left(\prod_{j} [Z^{(m)}_{j}]^{-1/2}\right)[u_i +\delta u_i]$, where the bare
dimensionless couplings are given by $u_{i,0}=N_0\,U_{i,0}$ (where $N_0$ is the density of states at the Fermi energy of the system), $u_i$ are the renormalized dimensionless couplings and $\delta u_i$ are the counterterms
that regularize the model order by order within perturbation theory. The prefactor $Z_j^{(m)}$ (where $m=\{d_1,f_1,f_2\}$ and $j=\{d_{xz},d_{yz}\}$) is the field renormalization strength of the effective theory that measures the low-energy coherence of the quasiparticle excitations
at the hole and electron pockets associated with the orbitals $d_{xz}$ and $d_{yz}$ in the present model.

\section{One-loop RG Approach}

In this section, we apply the RG method in order to investigate the behavior of the multiband model defined in Eqs. (\ref{Eq_01})-(\ref{Eq_03}) at lower energies, which in our present case will
signify lower temperatures.
As a first step towards this goal, we will compute the RG flow equations within the one-loop approximation in order to get a qualitative idea of the RG analysis when applied
to the present model. The one-loop RG approximation is equivalent to the so-called parquet approach originally developed by the Soviet school long time ago \cite{Bychkov}. As a consequence, we will now calculate
both the RG flow equations for the effective couplings in the present model and of some relevant order-parameter susceptibilities at this order within the perturbative RG scheme. A schematic representation of some Feynman diagrams of the vertex corrections calculated in the present work is shown in Fig. 2. We will postpone the two-loop RG calculation for the present model to a future section in this paper.

Since both hole and electron pockets of the FeSe compound are small, the one-loop RG flow equations for the effective dimensionless couplings describing the process in which the high-energy degrees of freedom are progressively integrated out in the model are given by 

\vspace{-0.1cm}

\begin{align}\label{Eq_04}
&\partial_l u_1=u_1^2+u_3^2,\nonumber\\
&\partial_l \bar{u}_1=\bar{u}_1^2+\bar{u}_3^2,\nonumber\\
&\partial_l u_2=2u_1u_2-2u_2^2,\nonumber\\
&\partial_l \bar{u}_2=2\bar{u}_1\bar{u}_2-2\bar{u}_2^2,\nonumber\\
&\partial_l u_3=-u_3 u_4+4u_3 u_1-u^{(1)}_5 u_3-\bar{u}_5 \bar{u}_3-2{u}_2 u_3,\nonumber\\
&\partial_l \bar{u}_3=-\bar{u}_3 u_4+4\bar{u}_3 \bar{u}_1-{u}^{(2)}_5\bar{u}_3 -\bar{u}_5 u_3-2\bar{u}_2 \bar{u}_3,\nonumber\\
&\partial_l u_4=-u_4^2-u_3^2-\bar{u}_3^2,\nonumber\\
&\partial_l u^{(1)}_5=-[u^{(1)}_5]^2-\bar{u}_5^2-u_3^2,\nonumber\\
&\partial_l u^{(2)}_5=-[u^{(2)}_5]^2-\bar{u}_5^2-\bar{u}_3^2,\nonumber\\
&\partial_l \bar{u}_5=-u^{(1)}_5\bar{u}_5-u^{(2)}_5\bar{u}_5-u_{3}\bar{u}_3,\nonumber\\
&\partial_l {\tilde{u}}_5=-(\tilde{u}_5)^{2}-(\tilde{\tilde{u}}_5)^{2},\nonumber\\
&\partial_l \tilde{\tilde{u}}_5=-2\tilde{u}_5 \tilde{\tilde{u}}_5,\nonumber\\
\end{align}

\noindent where $l=\frac{1}{2}\ln(\Lambda_0/\Lambda)$ is the RG step with $\Lambda_0$ being a fixed microscopic scale and $\Lambda$ being the floating RG scale. We note that in the above equations the RG scale must necessarily satisfy $\Lambda_0>\Lambda>E_F$, where $E_F$ is the Fermi energy of the system. In agreement with other works (see, e.g., Refs. \cite{Chubukov,Chubukov2,Honerkamp}), most couplings in the one-loop RG flow equations diverge at a critical
scale given by $l_c^{(1)}=\frac{1}{2}\ln(\Lambda_0/\Lambda_c^{(1)})$. In order to solve the above equations analytically, one could in principle make the following ansatz $u_i(l)=C_i/(l^{(1)}_c-l)$, with constant coefficients defined by $C_i$. In this way, at one-loop RG order, 
the ratio of the couplings would reach fixed values at very low energies and, consequently, display some degree of ``universality'' regardless of the initial conditions for the couplings. However, we 
point out here that this ``universality" is an artifact of the one-loop RG approximation and, as we will see later on, it breaks down completely at two-loop RG order. This ansatz procedure has been performed
in many works in the literature (see, e.g., Refs. \cite{Chubukov,Chubukov2}), but we will not do this in the present work. The reason for this choice is that the above scaling behavior only works, strictly speaking, in the very end of the one-loop RG flow, and
we believe it is better to track the flow from the high-energy (microscopic) scale to not-too-low energy scales. As a result, we will choose to solve all the coupled differential equations obtained in this work along the entire RG flow only by numerical means.

To investigate the precise nature of the enhanced correlations and the competing orders that appear at low energies in this minimal multiband model, we must add to Eqs. (\ref{Eq_01})-(\ref{Eq_03}) the following external fields coupled to
bilinear combinations of fermions, i.e.,

\begin{align}\label{Eq_06}
H_{ext}&=\sum_{\mathbf{k}}\Delta^{(d_1) \uparrow\downarrow}_{B,SC}(\mathbf{k,q}) d_{1,\mathbf{k+q/2},\uparrow}^{\dagger}d_{1,\mathbf{-k+q/2},\downarrow}^{\dagger}\nonumber\\
&+\sum_{\mathbf{k},i=1,2}\Delta^{(f_i) \uparrow\downarrow}_{B,SC}(\mathbf{k,q}) f_{i,\mathbf{k+q/2},\uparrow}^{\dagger}f_{i,\mathbf{-k+q/2},\downarrow}^{\dagger}\nonumber\\
&+\sum_{{\mathbf{k},\sigma}}\Delta^{(1) \sigma\sigma}_{B,DW}(\mathbf{k,q}) d_{1,\mathbf{k+q/2},\sigma}^{\dagger}f_{1,\mathbf{k-q/2},\sigma}+H.c.\nonumber\\
&+\sum_{{\mathbf{k},\sigma}}\Delta^{(2) \sigma\sigma}_{B,DW}(\mathbf{k,q}) d_{1,\mathbf{k+q/2},\sigma}^{\dagger}f_{2,\mathbf{k-q/2},\sigma}+H.c.\nonumber\\
&+\sum_{{\mathbf{k},\sigma}}\Delta^{(d_1) \sigma\sigma}_{B,DW}(\mathbf{k,q=0}) d_{1,\mathbf{k+q/2},\sigma}^{\dagger}d_{1,\mathbf{k-q/2},\sigma}\nonumber\\
&+\sum_{\substack{{\mathbf{k},\sigma}\\{i=1,2}}}\Delta^{(f_i) \sigma\sigma}_{B,DW}(\mathbf{k,q=0}) f_{i,\mathbf{k+q/2},\sigma}^{\dagger}f_{i,\mathbf{k-q/2},\sigma},
\end{align}

\noindent where $\Delta^{(d_1) \uparrow\downarrow}_{B,SC}(\mathbf{k,q})$ and $\Delta^{(d_1) \sigma\sigma}_{B,DW}(\mathbf{k,q})$ refer to the bare response vertices
associated with the superconducting (SC) and density-wave (DW) orders for the excitations near the hole pockets described by the operator $d_1$, and $\Delta^{(f_i) \uparrow\downarrow}_{B,SC}(\mathbf{k,q})$ and $\Delta^{(f_i) \sigma\sigma}_{B,DW}(\mathbf{k,q})$ stand for the bare response vertices
associated with the superconducting (SC) and density-wave (DW) orders for the excitations close to the electron pockets described by the operators $f_i$ ($i=1,2)$. These
added terms will generate new three-legged vertex corrections in the model (see Fig. 3) that turn out to be divergent in the low-energy limit. In a similar way as before, we renormalize the perturbation theory by defining
$\Delta_{B,l}^{(m)}=\left(\prod_{j} [Z^{(m)}_{j}]^{-1/2}\right)[\Delta_{R,l}^{(m)} +\delta \Delta_{R,l}^{(m)}]$, where $\Delta_{R,l}^{(m)}$ (for $l=DW$ or $SC$, $m=\{d_1,f_1,f_2\}$ and $j=\{d_{xz},d_{yz}\}$) is the corresponding order-parameter response vertices with $\delta \Delta_{R,l}^{(m)}$ being
the necessary counterterm to regularize the effective model.

Therefore, using standard RG prescriptions \cite{Freire,Freire2,Freire4}, and
by means of either a symmetrization or antisymmetrization procedure of the above response vertices with respect to the spin indices, we get the following order parameters

\begin{center}
$\left\{%
\begin{array}{ll}
     \Delta_{SDW}(\mathbf{Q_{x}})=\Delta_{R,DW}^{(1)\uparrow\uparrow}(\mathbf{q=Q_{x}})-\Delta_{R,DW}^{(1)\downarrow\downarrow}(\mathbf{q=Q_{x}}),\\
     \Delta_{SDW}(\mathbf{Q_{y}})=\Delta_{R,DW}^{(2)\uparrow\uparrow}(\mathbf{q=Q_{y}})-\Delta_{R,DW}^{(2)\downarrow\downarrow}(\mathbf{q=Q_{y}}),\\
     \Delta^{(m)}_{SSC}=\Delta_{R,SC}^{(m)\uparrow\downarrow}(\mathbf{q}=0)-\Delta_{R,SC}^{(m)\downarrow\uparrow}(\mathbf{q}=0),\\
     \Delta^{(m)}_{UCS}=\Delta_{R,DW}^{(m)\uparrow\uparrow}(\mathbf{q}=0)+\Delta_{R,DW}^{(m)\downarrow\downarrow}(\mathbf{q}=0),
\end{array}%
\right.$ 
\end{center}

\noindent where $SDW$, $UCS$, and $SSC$ stand for, respectively, to spin density wave, uniform charge susceptibility, 
and singlet superconducting order parameters.  In addition, $\Delta_{SDW}(\mathbf{Q_{x}})$ and $\Delta_{SDW}(\mathbf{Q_{y}})$ refer to stripe-type SDW magnetism with a modulation described, respectively, by the wavevectors $\mathbf{Q_{x}}=(\pi,0)$ and $\mathbf{Q_{y}}=(0,\pi)$.

Now, we can calculate the RG flow equations for the order parameters up to one-loop order in a standard way. A schematic representation of the Feynman diagrams for this calculation is given by Fig. 3. As a result, some of these RG equations are given by

\vspace{-0.3cm}

\begin{align}\label{Eq_04}
&\partial_l \Delta_{SDW}(\mathbf{Q_{x}})=(u_1+u_3)\Delta_{SDW}(\mathbf{Q_{x}}),\nonumber\\
&\partial_l \Delta_{SDW}(\mathbf{Q_{y}})=(\bar{u}_1+\bar{u}_3)\Delta_{SDW}(\mathbf{Q_{y}}),\nonumber\\
&\partial_l \Delta_{SSC}^{(d_1)}=(u_3+\bar{u}_3-u_4)\Delta_{SSC}^{(d_1)},\nonumber\\
&\partial_l \Delta_{SSC}^{(f_1)}=(u_3+\bar{u}_3-u^{(1)}_5)\Delta_{SSC}^{(f_1)},\nonumber\\
&\partial_l \Delta_{SSC}^{(f_2)}=(u_3+\bar{u}_3-u^{(2)}_5)\Delta_{SSC}^{(f_2)}.
\end{align}

\noindent Lastly, we can define the symmetries of some relevant order parameters in the following way

\begin{center}
$\left\{%
\begin{array}{ll}
     \Delta^{s_{++}}_{SSC}=\Delta_{SSC}^{(d_1)}+\Delta_{SSC}^{(f_1)}+\Delta_{SSC}^{(f_2)},\\
     \Delta^{s_{\pm}}_{SSC}=\Delta_{SSC}^{(d_1)}-\Delta_{SSC}^{(f_1)}-\Delta_{SSC}^{(f_2)},\\
     \Delta^{s_{\pm}}_{POM}=\Delta_{UCS}^{(d_1)}-\Delta_{UCS}^{(f_1)}-\Delta_{UCS}^{(f_2)},\\
     \Delta^{d_{\pm}}_{POM}=\Delta_{UCS}^{(f_2)}-\Delta_{UCS}^{(f_1)},
\end{array}%
\right.$ 
\end{center}

\noindent where $s_{++}$-SSC and $s_{\pm}$-SSC refer to, respectively, singlet superconducting order without any sign-change and singlet superconducting order with a sign-change in the pairing gap between electron and hole pockets in the model and the subscript POM stands for a Pomeranchuk instability (i.e., a $\mathbf{q}=0$ order that results in a spontaneous deformation of the Fermi surface of the system) that can in principle appear as a competing order in the present multiband model
in both $s_{\pm}$ and $d_{\pm}$ channels. 

\section{Self-energy effects and Uniform susceptibilities at two-loop RG order}

At one-loop RG order, it can be easily shown that there are no self-energy diagrams that yield non-analyticities as a function of the external energy $\omega$. In other
words, at this RG order, the quasiparticle weights associated with different orbital components on the Fermi surface do not renormalize as a function of the RG scale and their value will always remain equal to unity (i.e., the noninteracting value). In order to investigate 
possible orbital-selective renormalization of the quasiparticle weights, one must go at least up to two-loop RG order or beyond. This analysis will be performed now.

\begin{figure}[t]
\includegraphics[width=3.44in]{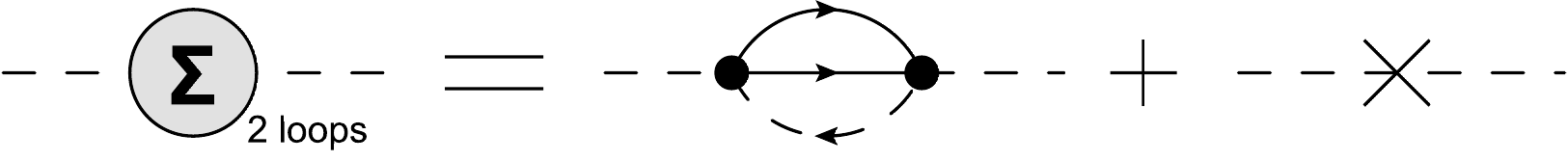}
\caption{Schematic representation of the non-analytic contribution to the self-energy of the present model. The diagram with a cross represents the corresponding counterterm, which regularizes the renormalized perturbation theory.}
\end{figure} 

A schematic representation of the two-loop self-energy diagram that displays non-analyticity as a function of the external energy is depicted in Fig. 4. Since both hole pockets and electron pockets are experimentally small in the FeSe compound,
we shall focus in this work on the regime in which the RG scale $\Lambda$ is larger than the Fermi energy (i.e., $\Lambda_0>\Lambda>E_F$). In this energy window, if we compute the self-energy
of the present multiband model, we obtain that its non-analytic contributions are given by sunset diagrams displayed schematically in Fig. 4, which approximately evaluate to

\vspace{-0.3cm}

\begin{align}
    \Sigma_{d_{yz}}(i\omega,\mathbf{k_F})&\approx \frac{{u}_3^2}{4}\,i\omega\ln\left(\frac{i\omega}{\Lambda_0}\right)+\cdots,\\
    \Sigma_{d_{xz}}(i\omega,\mathbf{k_F})&\approx \frac{\bar{u}_3^2}{4}\,i\omega\ln\left(\frac{i\omega}{\Lambda_0}\right)+\cdots.
\end{align}

\noindent Therefore, by following a standard RG procedure \cite{Peskin}, it can be shown that the quasiparticle weights of the present model are
described by the RG flow equations

\vspace{-0.3cm}

\begin{align}
   &\frac{\Lambda}{Z_{d_{yz}}}\frac{d Z_{d_{yz}}}{d\Lambda}=\frac{1}{4}{u}_3^2,\\    
   &\frac{\Lambda}{Z_{d_{xz}}}\frac{d Z_{d_{xz}}}{d\Lambda}=\frac{1}{4}\bar{u}_3^2,
\end{align}

\noindent where the initial conditions for these equations are given by $Z_{d_{xz}}(\Lambda_0)=Z_{d_{yz}}(\Lambda_0)=1$. We point out that we could in principle solve numerically the above two-loop equations together with the one-loop RG equations for the effective couplings.
However, this approximation turns out not to be consistent, since both flow equations must be evaluated at the same level in the RG theory. For this reason, it becomes of great importance to compute the full two-loop RG equations for the effective couplings in order to address if the orbitally-selective renormalization of the quasiparticle weight will appear or not in the present model. We will perform this full two-loop calculation in a subsequent section.

We digress for a moment and move on to the calculation of the uniform susceptibilities (\emph{i.e.} for $\mathbf{q}=0$) in the model. In this respect, we point out that Chubukov \emph{et al.} \cite{Chubukov} proceed to calculate the uniform susceptibilities of a similar itinerant model within a conventional perturbation theory at one-loop order with very interesting results. However, in Ref. \cite{Chubukov}, the equations describing the uniform susceptibilities obey algebraic rather than differential equations. In our view, it is better to compute all susceptibilities of the model on equal footing. This goal is achieved in the present work. But, unlike the calculation of the previous RG equations for the order parameters in which divergent Feynman diagrams that renormalize the corresponding response vertices already appear at one-loop order, the first singular diagrams only emerge at two-loop order in the computation of the uniform susceptibilities. These are the so-called Aslamazov-Larkin diagrams, which are schematically shown at the bottom part of Fig. 3. For this reason, we will now calculate these diagrams. By following the same approach as explained before to regularize the effective model at this order, we obtain that the corresponding two-loop RG flow equations for the uniform susceptibilities finally evaluate to

\vspace{-0.3cm}

\begin{align}\label{Eq_06}
&\partial_l 
\begin{bmatrix}
\Delta_{UCS}^{(d_1)}  \\ \Delta_{UCS}^{(f_1)} \\ \Delta_{UCS}^{(f_2)} \\
\end{bmatrix}=\hat{\mathbf{M}}\cdot \begin{bmatrix}
\Delta_{UCS}^{(d_1)}\\ \Delta_{UCS}^{(f_1)} \\ \Delta_{UCS}^{(f_2)} \\
\end{bmatrix},
\end{align}

\noindent with the matrix $\hat{\mathbf{M}}$ being given by

\vspace{-0.3cm}

\begin{align}
\hat{\mathbf{M}}=
\begin{bmatrix}
0 &  x  &  y \\
x  &  0 &  0 \\
y  &  0  &  0    \\
\end{bmatrix},
\end{align}

\noindent whose elements are given by $x=(u_1u_2-u_3^2-u_1^2-u_2^2)/2$ and $y=(\bar{u}_1\bar{u}_2-\bar{u}_3^2-\bar{u}_1^2-\bar{u}_2^2$)/2.

\section{Preliminary Numerical Results}

In this section, we will solve numerically all the RG equations obtained thus far in this work. The one-loop RG flow for the couplings are depicted in Fig. 5. Here, we choose the following parametrization $U=0.122$, $U'=0.082$, and $J=J'=0.02$ (in units of $N_0^{-1}$), which is motivated
by previous RG studies in the literature \cite{Chubukov,Honerkamp}. We emphasize here that we enforce the spin-rotational symmetry condition for this model, i.e., $J=(U-U')/2$. As a result, 
we find that many dimensionless couplings associated with interaction processes among the different orbitals scale differently under the RG flow with the Umklapp scattering described by $u_3$ becoming the leading effective coupling at this order. As can be seen in Fig. 5, many dimensionless couplings in the system clearly flow to strong coupling (which, in our case, means that they scale to numerical values larger than one), while a few other couplings have the tendency to scale to zero at low energies. These latter couplings will therefore become irrelevant in the low-energy effective description.

\begin{figure}[t]
\includegraphics[width=3.7in]{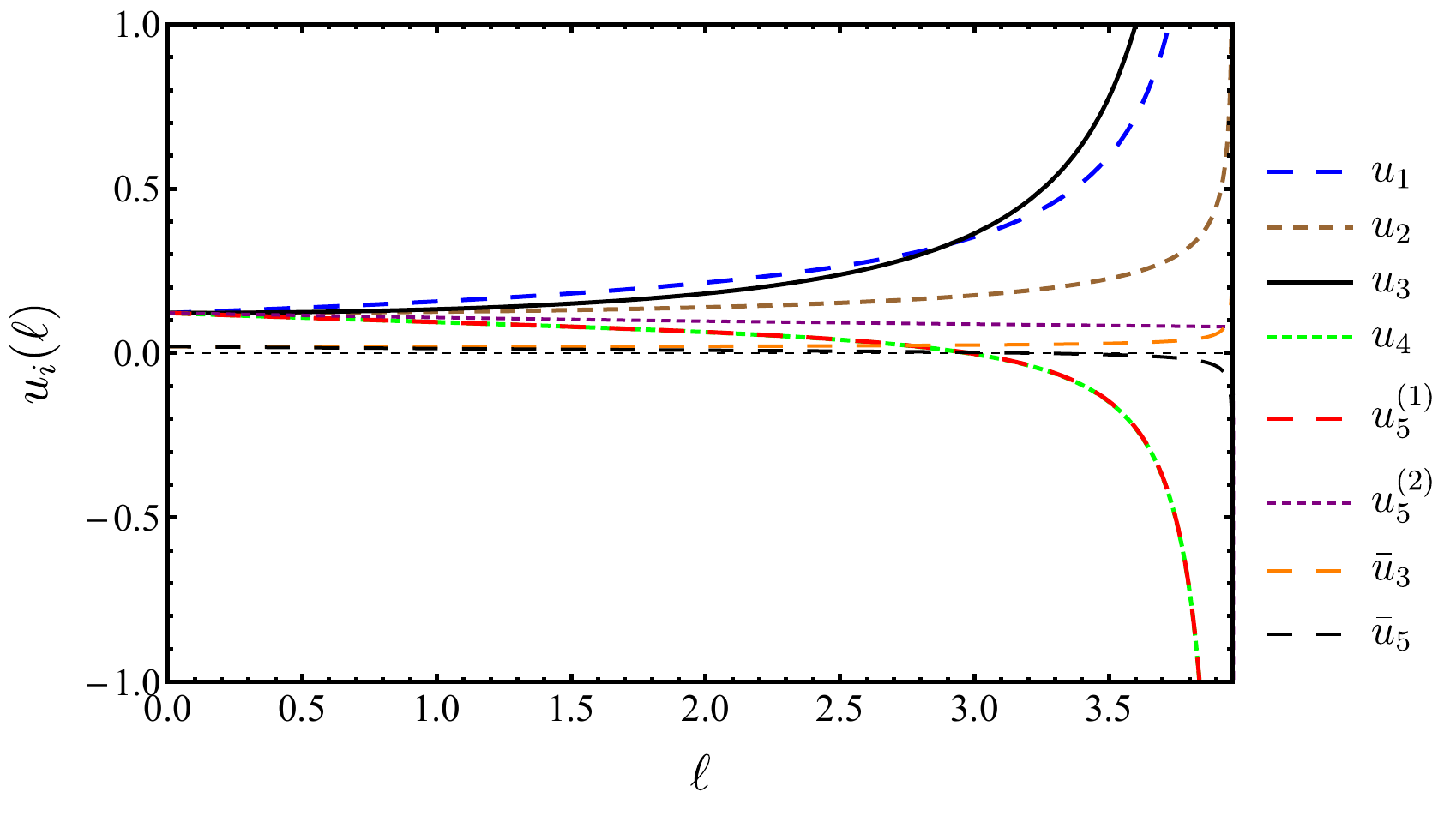}
\caption{(Color online) The one-loop RG flow for some relevant couplings $u_i$ as a function of the RG step $l$. We choose the parametrization $U=0.122$, $U'=0.082$, and $J=J'=0.02$ (in units of $N_0^{-1}$).}
\end{figure}

By solving numerically also the RG equations for the competing order-parameter response vertices, we obtain that the one-loop RG flows are displayed in Fig. 6 (for the specific case of the uniform susceptibilities, we note that we consider here, as a first estimate, the one-loop RG equations for the couplings and the two-loop RG equations for the uniform response vertices). In this plot, we
find that the leading instability is given by a Pomeranchuk instability in the $s_{\pm}$-channel, followed closely by a Pomeranchuk instability in the $d_{\pm}$-channel, then stripe-type antiferromagnetic $SDW(\mathbf{Q_x})$ fluctuations in the third place and, fourthly, by the
$s_{\pm}$ singlet pairing tendencies (we point out that the $s_{++}$ superconducting correlations only appear in fifth place and is therefore subleading to $s_{\pm}$ pairing). Taken at face value, the second leading Pomeranchuk instability in the $d_{\pm}$ channel would suggest the formation of a nematic order in the presence of pronounced short-range (stripe-type) antiferromagnetic correlations in the system.
This result agrees qualitatively with the results obtained by Chubukov \emph{et al.} \cite{Chubukov}, even though our technical details differ from theirs. As a consequence, they interpreted their results in terms of a scenario in which the source
of nematicity in the FeSe compound could be attributed to an orbital order induced by a Pomeranchuk instability in the model. 
This order implies the breaking of the degeneracy of the on-site energies of the $d_{xz}$ and $d_{yz}$, which leads to an spontaneous deformation (elongation in one direction and compression in the other direction) of the Fermi pockets without translation symmetry breaking in the system. Changing the doping concentration or by applying external pressure in the model would lead to a rapid destruction of the approximate nesting condition of the Fermi pockets, which would cease the contributions of both the Pomeranchuk-type and the antiferromagnetic response vertices. As a result, the next-to-leading $s_{\pm}$ pairing instability would naturally set in. This scenario would in principle agree with the experimental data in the FeSe compound, where it is indeed observed a nematic transition from tetragonal order to orthorhombic symmetry with no accompanying magnetic order (but with significant structure in the spin fluctuations) and an adjacent unconventional superconductivity in the corresponding phase diagram.

\begin{figure}[t]
\includegraphics[width=3.2in]{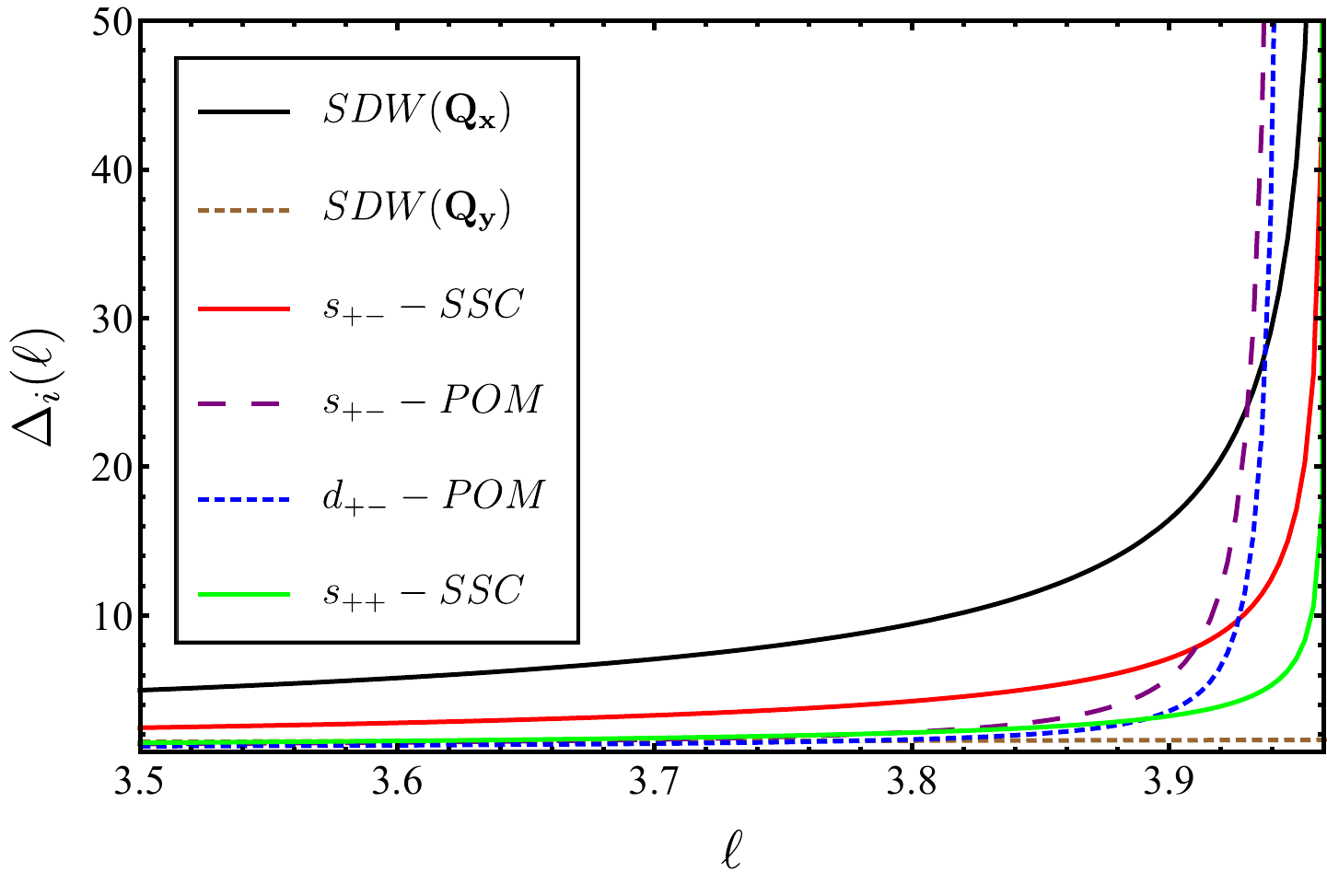}
\caption{(Color online) One-loop RG flow for some relevant order-parameter susceptibilities as a function of the RG step $l$. The initial conditions for these susceptibilities are chosen to be $\Delta_i(l=0)=1$. For the initial couplings, we choose the following parametrization $U=0.122$, $U'=0.082$, and $J=J'=0.02$ (in units of $N_0^{-1}$).}
\end{figure}

However, as advertised previously in our work, pertinent questions arise at this point: How robust is the one-loop result with respect to including higher-order vertex corrections
and nontrivial self-energy effects at two-loop RG order or beyond? In other words, what is the main effect of the self-energy feedback onto the corresponding RG flow equations in the present model?
Since, as we have seen before, the one-loop RG theory is ultimately not controlled analytically at low energies, this turns out to be a central question that we will try to make the first steps to provide an answer in the next section.

\section{Two-loop RG calculation}

Now, we move on to the implementation of the two-loop RG framework to the present multiband model in order to describe qualitatively the FeSe compound within the regime such that $\Lambda_0>\Lambda>E_F$. As we have mentioned earlier, this RG analysis goes beyond the parquet approximation and was explained by one of the authors in the context of different models in other works \cite{Freire,Freire2,Freire3}. From the point of view of number of diagrams taken into account, the two-loop RG approach turns out to be much more involved than the one-loop RG scheme. To emphasize this point, a schematic representation of some Feynman diagrams of the vertex corrections calculated in the present work is shown in Fig. 2. 
In addition, we point out here that, since both $\tilde{u}_5$ and $\tilde{\tilde{u}}_5$ interactions flow to zero in the low-energy limit within the one-loop RG scheme, we will neglect, for simplicity, only those couplings at two-loop RG order. The corresponding two-loop RG equations associated with the present model are given explicitly in Appendix A.

The numerical solution of the two-loop RG flow equations is displayed in Fig. 7. In this plot, we observe that the effective couplings continue to diverge at
a finite energy scale $\Lambda^{(2)}_c=\Lambda_0 e^{-2l^{(2)}_c}$, which turns out to be similar to the energy scale that the one-loop RG flow diverges obtained previously in this work (i.e., $\Lambda^{(2)}_c\approx \Lambda^{(1)}_c$). Despite that, the tendencies of the RG flow of some interactions
change qualitatively as compared to the one-loop case with, e.g., the couplings $u_2$, $\bar{u}_3$ and $\bar{u}_5$ being now suppressed in the low-energy limit, while the $u_3$ scattering process continues to be the dominant coupling (albeit with a slight quantitative variation) in the effective model.

\begin{figure}[t]
\includegraphics[width=3.7in]{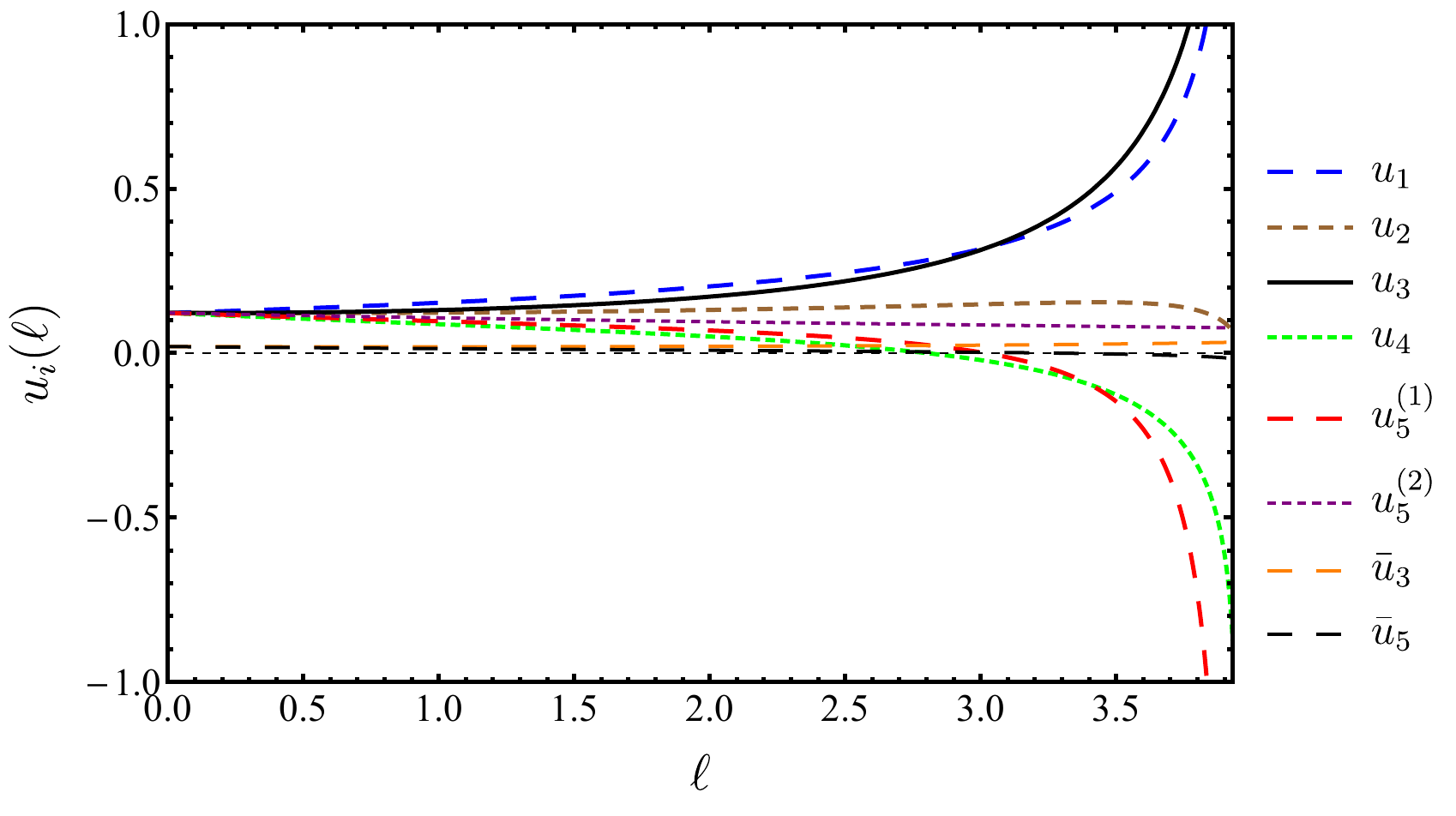}
\caption{(Color online) The two-loop RG flow for some relevant couplings $u_i$ as a function of the RG step $l$. We choose the same parametrization as before, i.e., $U=0.122$, $U'=0.082$, and $J=J'=0.02$ (in units of $N_0^{-1}$).}
\end{figure}

We then proceed to the analysis of the RG flow of the quasiparticle weight at two-loop order (shown in Fig. 8). In this plot, we see that
orbitally-selective renormalization interestingly shows up with the $Z_{d_{yz}}$-factor associated with the $d_{yz}$ orbital being more suppressed than
$Z_{d_{xz}}$ related to the $d_{xz}$ orbital. Despite that, we observe that both of them remain finite at low energies and, as a consequence, the Landau quasiparticle excitations indeed remain coherent at these scales. It is reasonable to speculate that, as the (initial) microscopic interactions become stronger, this tendency towards orbital-selectivity could become even more pronounced.
This may suggest that for the physical situation that applies to some iron-based chalcogenides (where the microscopic interactions are expected to be within an intermediate coupling regime), this property of the system could become even more
enhanced. As a consequence, our present results would imply qualitatively a scenario that exhibits a hierarchy between the quasiparticle weights associated with different orbitals in the following way $Z_{d_{xy}}\ll Z_{d_{yz}}< Z_{d_{xz}}$. 
This hierarchy has some qualitative differences compared to a recent phenomenology proposed in Ref. \cite{Hirschfeld}, which describes the modulations of the superconducting gap in FeSe by assuming a spin-fluctuation-mediated pairing mechanism. This fact suggests that further extensions of the present multiband model (e.g., by including more orbitals from the outset or including more pockets in the effective theory) are necessary in order to describe quantitatively the experimental data of these compounds. We plan to do this in a future work.

Next, we calculate the RG flow equations for some relevant competing orders within the two-loop RG approximation by following the procedure explained previously in this work. As a result, we obtain that 

\vspace{-0.3cm}

\begin{align}\label{Eq_04}
&\partial_l \Delta_{SDW}(\mathbf{Q_{x}})=\left[u_1+u_3-\frac{1}{4}\left(2u_3^2+\bar{u}_3^2\right)\right]\Delta_{SDW}(\mathbf{Q_{x}}),\nonumber\\
&\partial_l \Delta_{SDW}(\mathbf{Q_{y}})=\left[\bar{u}_1+\bar{u}_3-\frac{1}{4}\left(u_3^2+2\bar{u}_3^2\right)\right]\Delta_{SDW}(\mathbf{Q_{y}}),\nonumber\\
&\partial_l \Delta_{SSC}^{(d_1)}=\left[u_3+\bar{u}_3-u_4-\frac{1}{2}\left(u_3^2+\bar{u}_3^2\right)\right]\Delta_{SSC}^{(d_1)},\nonumber\\
&\partial_l \Delta_{SSC}^{(f_1)}=\left(u_3+\bar{u}_3-u^{(1)}_5-\frac{1}{2}u_3^2\right)\Delta_{SSC}^{(f_1)},\nonumber\\
&\partial_l \Delta_{SSC}^{(f_2)}=\left(u_3+\bar{u}_3-u^{(2)}_5-\frac{1}{2}\bar{u}_3^2\right)\Delta_{SSC}^{(f_2)}.
\end{align}

The numerical solution of the two-loop RG flow of the order-parameter response vertices is depicted in Fig. 9.
From this plot, we observe that, due to the self-energy feedback onto the equations, the RG flow of the order-parameter susceptibilities changes not only quantitatively but also qualitatively. The leading susceptibility now becomes the stripe-type antiferromagnetic $SDW(\mathbf{Q_x})$, indicating the dominance of short-range magnetic spin fluctuations in the system. The second leading instability at two-loop RG order turns out to be the $s_{\pm}$ superconducting susceptibility, followed closely by $s_{++}$ pairing tendencies. As expected, this confirms that stripe-type antiferromagnetic $SDW(\mathbf{Q_x})$ fluctuations indeed promote the leading $s_{\pm}$ pairing instability in the system \cite{Mazin,Chubukov6}. Despite that, due to strong orbital fluctuations also present in the model, superconductivity in the $s_{++}$ channel is a close second pairing instability \cite{Kontani}.
Meanwhile, as can be seen in Fig. 9, the renormalization of the Pomeranchuk vertices in both $s_{\pm}$ and $d_{\pm}$ channels become softened by the self-energy effects and change its relative position to fourth and fifth potential instabilities in the system, respectively.
This suggests that the Pomeranchuk instability in the $d_{\pm}$ channel may not be the underlying mechanism of nematicity in this itinerant multiband model. In fact, from the previous result regarding the quasiparticle weight,
we may conclude that another possible scenario is that orbitally-selective renormalization of the quasiparticle weights associated with different orbitals may be the origin of nematicity in the present model. This orbital selectivity is induced by the presence of short-range stripe-type antiferromagnetic fluctuations in the system. This conclusion shares qualitative features with
very recent STM data \cite{Sprau} together with complementary phenomenological calculations \cite{Hirschfeld}, which emphasize that the origin of the $C_4$ symmetry breaking down to $C_2$ in the FeSe compound can be indeed attributed to an anisotropic renormalization of the quasiparticle weights at different orbitals as result of the spin fluctuations. 

\begin{figure}[t]
\includegraphics[width=3.25in]{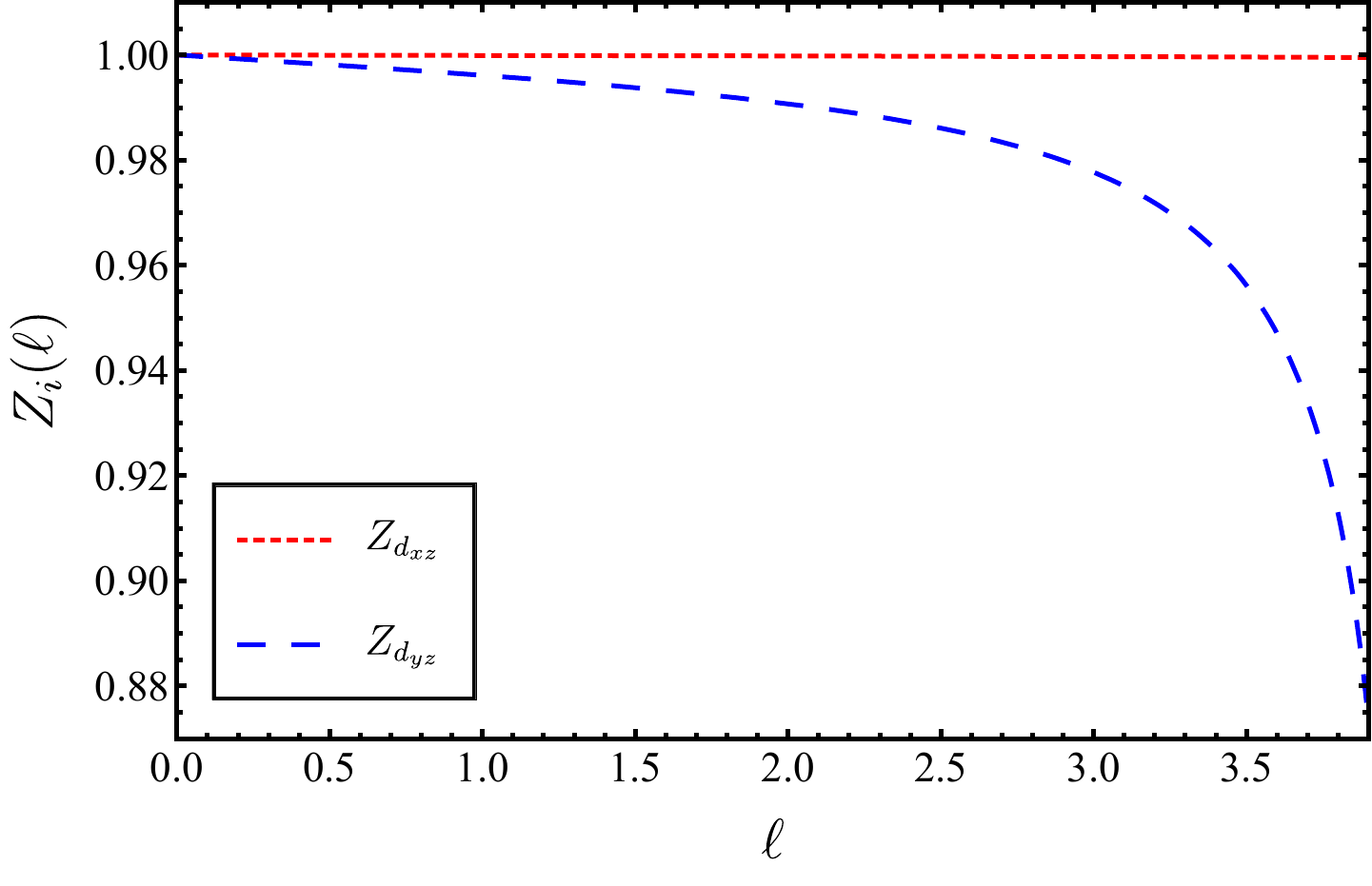}
\caption{(Color online) Two-loop RG flow for the quasiparticle weight $Z_i$ as a function of the RG step $l$. For the initial couplings, we choose the same parametrization as before, i.e., $U=0.122$, $U'=0.082$, and $J=J'=0.02$ (in units of $N_0^{-1}$).}
\end{figure}

\begin{figure}[t]
\includegraphics[width=3.2in]{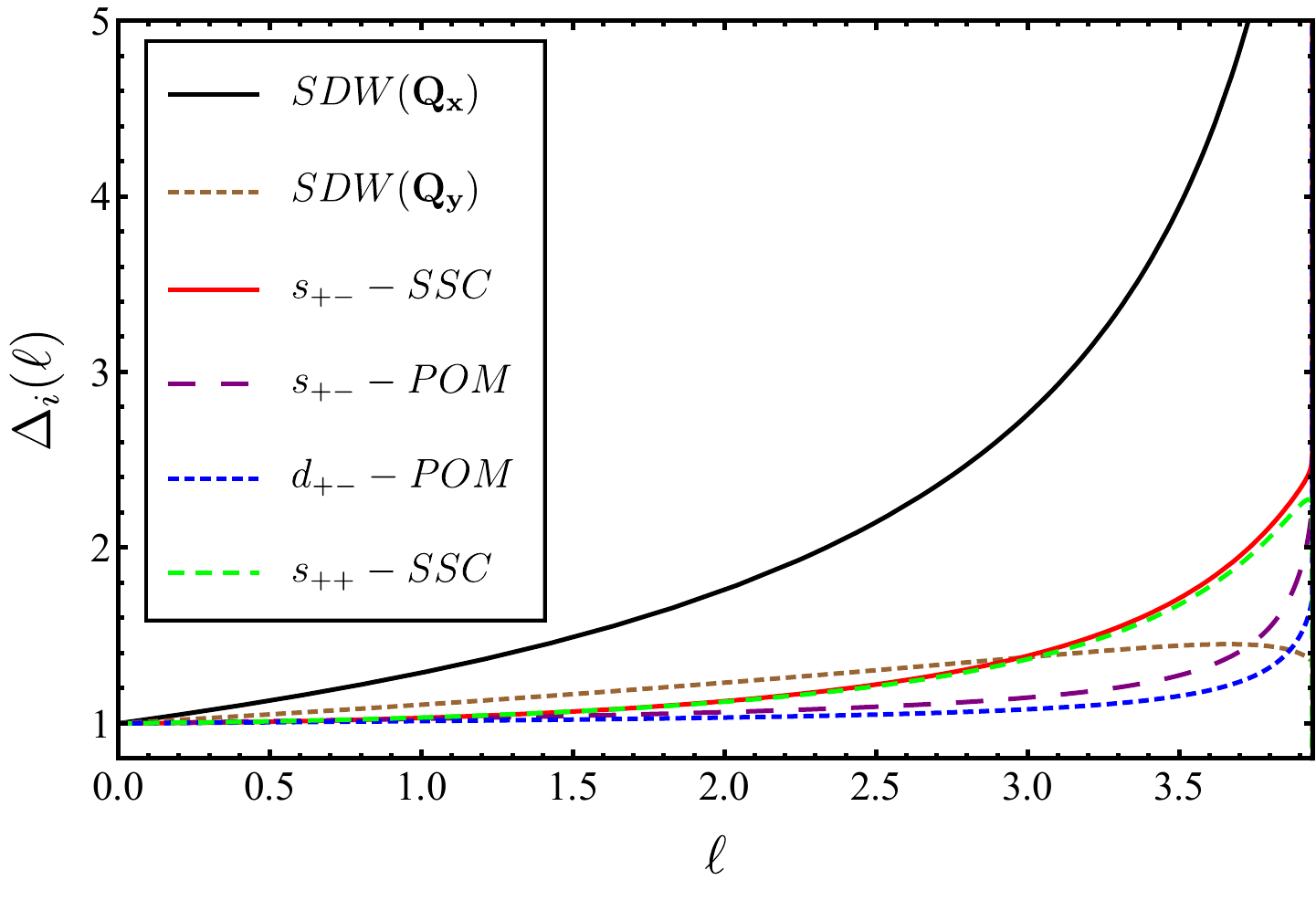}
\caption{(Color online) Two-loop RG flow for the order-parameter susceptibilities as a function of the RG step $l$. The initial conditions for these susceptibilities are chosen to be $\Delta_i(l=0)=1$. For the initial couplings, we choose the same parametrization as before, i.e., $U=0.122$, $U'=0.082$, and $J=J'=0.02$ (in units of $N_0^{-1}$).}
\end{figure}

Moreover, we observe from Fig. 9 the corresponding effect of the two-loop quantum fluctuations on the dominant stripe-type antiferromagnetic $SDW(\mathbf{Q_x})$ vertex in the model. It can be clearly seen that its divergence is strongly reduced as compared to the one-loop case (see again Fig. 6). In this way, the two-loop contributions act to push the divergence of the $SDW(\mathbf{Q_x})$ susceptibility towards higher values of the RG step $l$ (i.e., to lower temperatures). This is an expected effect and it is related to the well-known Mermin-Wagner theorem, which states that no spontaneous breaking of a continuous symmetry is possible at finite temperatures in a 2D model. Therefore, it is plausible to expect that by including the quantum fluctuations to all orders within the present perturbative RG scheme will suppress the stripe-type antiferromagnetic onset temperature $T_N$ in the system down to zero.
This scenario may be the underlying reason as to why no antiferromagnetic order is observed in the FeSe compound at ambient pressure. In other words, our scenario implies a sort of ``magnetic frustration'' induced by the presence of strong quantum fluctuations in the system, which is qualitatively captured by our two-loop RG approach. 
Therefore, we believe that the results obtained in the present work may provide a simple microscopic setting from a weak-to-moderate perspective, in which the fundamental role of orbital selectivity to describe the physical properties of some iron-chalcogenide systems is emphasized.

It is interesting to compare our present two-loop RG results with other theoretical works that provide different scenarios to describe the nematicity of the FeSe compound.
Our work shares common findings with Ref. \cite{Chubukov5}, where it was proposed that the source of nematicity in the FeSe has similarities to the case of Fe pnictides (i.e., Ising nematic order due to stripe-type antiferromagnetic fluctuations), although they did not consider the renormalization of the quasiparticle weight, which in our scenario turns out to be important, as we have seen before. In Ref. \cite{Chubukov5}, the large difference between nematic order ($T_s=90$ K) and the magnetic transition ($T_N=0$) was also attributed to the fact that the small pockets in FeSe make the quantum fluctuations stronger, so that they effectively suppress the antiferromagnetic order down to zero, which of course has some analogies with our present scenario. Another appealing work that also has some parallels with our present scenario is the Ref. \cite{Fanfarillo}, where these authors use an Eliashberg theory in the context of a different itinerant model (that includes spin-orbit interaction) in the presence of orbital-selective spin fluctuations to compute many effects associated with the orbital-dependent self-energy corrections. On the other hand, the possible role of magnetic frustration in the FeSe compound is most clearly emphasized in the strong-coupling works, e.g., of the Refs. \cite{Si2,Nevidomskyy,DHLee,Si3}, where these authors put forward different localized spin models to describe this system and attribute the origin of nematicity in FeSe to different emerging phases: an antiferroquadrupolar order \cite{Si2}, a spin ferroquadrupolar order \cite{Nevidomskyy} and a nematic quantum paramagnetic phase \cite{DHLee}. As explained before, our present two-loop RG work may suggest some possible limitations of the weak-to-moderate coupling scenario to describe quantitatively the FeSe compound. Consequently, other
complementary stronger coupling models may be ultimately needed in order to achieve quantitative agreement between the theoretical calculations and the experimental data observed in this system.

\section{Summary}

We have performed a two-loop renormalization group (RG) investigation of a 2D effective multiband model with orbital structure, which is relevant for
describing the low-energy properties of some iron-chalcogenide superconductors. This problem is central in the field of iron-based superconductivity and, for this reason, it attracted considerable interest recently from many groups. In this respect, new aspects of our present calculation consist in the consideration of higher-order contributions in the RG scheme
that go beyond the widely-used parquet approximation and the consequent inclusion of nontrivial self-energy effects of the model
that result in an anisotropic renormalization of the quasiparticle weight at different orbitals in the metallic phase of these systems. From this RG analysis, we conclude from a simple model that the underlying origin of nematicity in these compounds may come from orbital-selective quasiparticle renormalization at the Fermi pockets, instead of a Pomeranchuk instability in the $d_{\pm}$ channel. This orbital selectivity emerges due to the presence of short-range stripe-type antiferromagnetic fluctuations in the system. 
In this way, we believe that the present two-loop RG results may contribute, from a weak-to-moderate coupling perspective, to the ongoing debate in the literature regarding the microscopic mechanism of the nematic phase observed in these materials.

\section*{Acknowledgments}\label{Section_SV}

One of us (H.F.) acknowledges funding from CNPq under Grant No. 405584/2016-4. Partial support from the Fundação de Amparo à Pesquisa do Estado de Goiás (FAPEG) is also greatly appreciated. 


\appendix

\section*{Appendix A}  

At two-loop RG order, by calculating the higher-order vertex corrections (see Fig. 2) and by including the self-energy feedback into the equations, the corresponding RG flows for the effective couplings in the model become described by the
following coupled differential equations

\begin{widetext}

\begin{align}\label{Eq_04}
\partial_l u_1&=u_1^2+u_3^2+\frac{1}{2}\big[-u_4u_1^2-u_5^{(1)}u_1^2-u_5^{(1)}u_3^2-u_4u_3^2\nonumber\\
&+2u_1u_3^2+u_1\bar{u}_3^2-2u_4u_2^2-2u_5^{(1)}u_2^2+2u_1u_2u_4\nonumber\\
&+2u_1u_2u_5^{(1)}-u_1(2u_3^2+\bar{u}_3^2)\big],\nonumber\\
\partial_l \bar{u}_1&=\bar{u}_1^2+\bar{u}_3^2+\frac{1}{2}\big[-u_4\bar{u}_1^2-u_5^{(2)}\bar{u}_1^2-u_5^{(2)}\bar{u}_3^2-u_4\bar{u}_3^2\nonumber\\
&+2\bar{u}_1\bar{u}_3^2+\bar{u}_1{u}_3^2-2u_4\bar{u}_2^2-2{u}_5^{(2)}\bar{u}_2^2+2\bar{u}_1\bar{u}_2{u}_4\nonumber\\
&+2\bar{u}_1\bar{u}_2{u}_5^{(2)}-\bar{u}_1(u_3^2+2\bar{u}_3^2)\big],\nonumber\\
\partial_l u_2&=2u_1u_2-2u_2^2+\frac{1}{2}\big[-2u_5^{(1)}u_2^2-2u_4u_2^2+2u_5^{(1)}u_1u_2\nonumber\\
&+2u_4u_1u_2-u_2\bar{u}_3^2-2u_2{u}_3^2-u_2(2u_3^2+\bar{u}_3^2)\big],\nonumber\\
\partial_l \bar{u}_2&=2\bar{u}_1\bar{u}_2-2\bar{u}_2^2+\frac{1}{2}\big[-2u_5^{(2)}\bar{u}_2^2-2u_4\bar{u}_2^2+2u_5^{(2)}\bar{u}_1\bar{u}_2\nonumber\\
&+2u_4\bar{u}_1\bar{u}_2-\bar{u}_2{u}_3^2-2\bar{u}_2\bar{u}_3^2-\bar{u}_2(u_3^2+2\bar{u}_3^2)\big],\nonumber\\
\partial_l u_3&=-u_3 u_4+4u_3 u_1-u_5^{(1)} u_3-\bar{u}_5 \bar{u}_3-2{u}_2{u}_3-\frac{1}{2}u_3(2u_3^2+\bar{u}_3^2),\nonumber\\
\partial_l \bar{u}_3&=-\bar{u}_3 u_4+4\bar{u}_3 \bar{u}_1-u_5^{(2)}\bar{u}_3-\bar{u}_5 u_3-2\bar{u}_2\bar{u}_3-\frac{1}{2}\bar{u}_3(u_3^2+2\bar{u}_3^2),\nonumber\\
\partial_l u_4&=-u_4^2-u_3^2-\bar{u}_3^2-2u_1^3-2\bar{u}_1^3-2u_2^3\nonumber\\
&-2\bar{u}_2^3+u_2u_1^2+\bar{u}_2\bar{u}_1^2+2{u}_2{u}_3^2+2\bar{u}_2\bar{u}_3^2\nonumber\\
&-2u_4(u_3^2+\bar{u}_3^2),\nonumber\\
\partial_l u_5^{(1)}&=-[u_5^{(1)}]^2-\bar{u}_5^2-{u}_3^2-2u_1 u_3^2-2u_1^3\nonumber\\
&+3u_2 u_1^2+3u_2 u_3^2-2u_5^{(1)}u_3^2,\nonumber\\
\partial_l u_5^{(2)}&=-[u_5^{(2)}]^2-\bar{u}_5^2-\bar{u}_3^2-2\bar{u}_1 \bar{u}_3^2-2\bar{u}_1^3\nonumber\\
&+3\bar{u}_2 \bar{u}_1^2+3\bar{u}_2 \bar{u}_3^2-2u_5^{(2)}\bar{u}_3^2,\nonumber\\
\partial_l \bar{u}_5&=-u_5^{(1)}\bar{u}_5-u_5^{(2)}\bar{u}_5-u_3\bar{u}_3-\frac{1}{2}\bar{u}_5(u_3^2+\bar{u}_3^2),\nonumber\\
\end{align}

\end{widetext}

\noindent where, as defined previously, $l=\frac{1}{2}\ln(\Lambda_0/\Lambda)$ is the RG step, $\Lambda_0$ is a fixed microscopic scale and $\Lambda$ is the floating RG scale. We note that in the above equations the RG scale must necessarily satisfy $\Lambda_0>\Lambda>E_F$.


\begin{thebibliography}{999}
\bibitem{Sprau} P. O. Sprau, A. Kostin, A. Kreisel, A. E. Böhmer, V. Taufour, P. C. Canfield, S. Mukherjee, P. J. Hirschfeld, B. M. Andersen, and J. C. Séamus Davis, Science \textbf{357}, 75 (2017).
\bibitem{Hirschfeld} A. Kreisel, B. M. Andersen, P. O. Sprau, A. Kostin, J.C. Séamus Davis, and P. J. Hirschfeld, Phys. Rev. B \textbf{95}, 174504 (2017).
\bibitem{Kostin} A. Kostin, P. O. Sprau, A. Kreisel, Yi Xue Chong, A. E. Böhmer, P. C. Canfield, P. J. Hirschfeld, B. M. Andersen and J. C. Séamus Davis, 
Nature Materials \textbf{17}, 869 (2018).
\bibitem{Kreisel} For a recent review on the FeSe superconductors, see, e.g., A. E. Böhmer and A. Kreisel, J. Phys. Condens. Matter \textbf{30}, 023001 (2017).
\bibitem{Hosono} Y. Kamihara, T. Watanabe, M. Hirano, and H. Hosono, J. Am. Chem. Soc. \textbf{130}, 3296 (2008).
\bibitem{Greene} J. Paglione and R. L. Greene, Nat. Phys. \textbf{6}, 645 (2010).
\bibitem{Jonhston} D. C. Johnston, Adv. Phys. \textbf{59}, 803 (2010).
\bibitem{Stewart} G. R. Stewart, Rev. Mod. Phys., \textbf{83}, 1589 (2011).
\bibitem{Cava} T. M. McQueen, A. J. Williams, P. W. Stephens, J. Tao, Y. Zhu, V. Ksenofontov, F. Casper,
C. Felser, and R. J. Cava, Phys. Rev. Lett., \textbf{103}, 057002 (2009).
\bibitem{Wu} F.-C. Hsu, J.-Y. Luo, K.-W. Yeh, T.-K. Chen, T.-W. Huang, P. M. Wu, Y. C.
Lee, Y.-L. Huang, Y.-Y. Chu, D.-C. Yan, and M.-K. Wu, Proc. Natl. Acad. Sci. U.S.A
\textbf{105}, 14262 (2008).
\bibitem{Sadovskii} M. V. Sadovskii, Physics-Uspekhi,
\textbf{59}, 947 (2016).
\bibitem{Wang} Z. Wang, C. Liu, Y. Liu, and J. Wang, J. Phys. Condens. Matter \textbf{29}, 153001 (2017).
\bibitem{Hoffman} D. Huang and J. E. Hoffman, Ann. Rev. Cond.
Matt. Phys., \textbf{8}, 311 (2017).
\bibitem{Jia} J.-F. Ge, Z.-L. Liu, C. Liu, C.-L. Gao, D. Qian, Q.-K. Xue, Y. Liu,
and J.-F. Jia,
Nat. Mater., \textbf{14}, 285 (2015).
\bibitem{Shi} X. Shi \emph{et al.}, Nat. Commun. \textbf{8}, 14988 (2017).
\bibitem{Shankar} R. Shankar, Rev. Mod. Phys. \textbf{66}, 129 (1994).
\bibitem{Freire} H. Freire, E. Correa and A. Ferraz,  Phys. Rev. B \textbf{71}, 165113 (2005); H. Freire, E. Corrêa and A. Ferraz, Phys. Rev. B \textbf{78}, 125114 (2008).
\bibitem{Freire2} V. S. de Carvalho and H. Freire, EPL (Europhysics Letters) \textbf{96}, 17006 (2011).
\bibitem{Chubukov} A. V. Chubukov, M. Khodas, and R. M. Fernandes, Phys. Rev. X \textbf{6}, 041045 (2016). 
\bibitem{Chubukov2} L. Classen, R.-Q. Xing, M. Khodas, and A. V. Chubukov, Phys. Rev. Lett. \textbf{118}, 037001 (2017).
\bibitem{Honerkamp} C. Honerkamp, Phys. Status Solidi B \textbf{254}, 1600151 (2017).
\bibitem{Matsuda} S. Kasahara, T. Watashige, T. Hanaguri, Y.i Kohsaka, T. Yamashita,
Y. Shimoyama, Y. Mizukami, R. Endo, H. Ikeda, K. Aoyama, T.
Terashima, S. Uji, T. Wolf, H. v. Löhneysen, T. Shibauchi, and Y.
Matsuda, Proc.
Natl. Acad. Sci. U.S.A  \textbf{111}, 16309 (2014).
\bibitem{Matsuda2} S. Kasahara, T. Yamashita, A. Shi, R. Kobayashi, Y. Shimoyama, T Watashige, K. Ishida,
T. Terashima, T. Wolf, F. Hardy, C. Meingast, H. v. Löhneysen, A. Levchenko, T. Shibauchi,
and Y. Matsuda, Nat. Commun. \textbf{7}, 12843 (2016).
\bibitem{Chubukov3} A. V. Chubukov, I. Eremin, and D. V. Efremov, Phys. Rev. B \textbf{93}, 174516 (2016).
\bibitem{Si} R. Yu, J.-X. Zhu and Q. Si, Phys. Rev. B \textbf{89}, 024509 (2014).
\bibitem{Kotliar} Z. P. Yin, K. Haule, and G. Kotliar, Nat. Mater. \textbf{10}, 932
(2011).
\bibitem{Georges} A. Georges, L. de Medici, and J. Mravlje, Annu. Rev. Condens. Matter Phys. \textbf{4}, 137 (2013). 
\bibitem{Suzuki} Y. Suzuki, T. Shimojima, T. Sonobe, A. Nakamura, M. Sakano, H. Tsuji, J. Omachi, K. Yoshioka, M. Kuwata-Gonokami, T. Watashige, R. Kobayashi, S. Kasahara, T. Shibauchi, Y. Matsuda, Y. Yamakawa, H. Kontani, and K. Ishizaka,
Phys. Rev. B \textbf{92}, 205117 (2015).
\bibitem{Coldea} M. D. Watson, T. K. Kim, L. C. Rhodes, M. Eschrig, M. Hoesch, A. A. Haghighirad, and A. I. Coldea,
Phys. Rev. B \textbf{94}, 201107(R) (2016).
\bibitem{Zhou} X. Liu, L. Zhao, S. He, J. He, D. Liu, D. Mou, B. Shen, Y. Hu,
J. Huang, and X. J. Zhou, J. Phys. Condens. Matter \textbf{27}, 183201 (2015).
\bibitem{Kordyuk} Y. V. Pustovit and A. A. Kordyuk, Low Temp. Phys. \textbf{42}, 995 (2016).
\bibitem{Terashima} T. Terashima, N. Kikugawa, A. Kiswandhi, E.-S. Choi, J. S. Brooks, S. Kasahara, T. Watashige, H. Ikeda, T. Shibauchi, Y. Matsuda, T. Wolf, A. E. Böhmer, F. Hardy, C. Meingast, H. v. Löhneysen, M.-T. Suzuki, R. Arita, and S. Uji,
Phys. Rev. B \textbf{90}, 144517 (2014).
\bibitem{Coldea2} M. D. Watson, T. Yamashita, S. Kasahara, W. Knafo, M. Nardone, J. Béard, F. Hardy, A. McCollam, A. Narayanan, S. F. Blake, T. Wolf, A. A. Haghighirad, C. Meingast, A. J. Schofield, H. v. Löhneysen, Y. Matsuda, A. I. Coldea, and T. Shibauchi, Phys. Rev. Lett. \textbf{115}, 027006 (2015).
\bibitem{Hanaguri} T. Hanaguri, K. Iwaya, Y. Kohsaka, T. Machida, T. Watashige, S. Kasahara, T. Shibauchi, Y. Matsuda, Sci. Adv. \textbf{4}, eaar6419 (2018).
\bibitem{Borisenko} S. V. Borisenko, D. V. Evtushinsky, Z.-H. Liu, I. Morozov, R. Kappenberger, S. Wurmehl,
B. Büchner, A. N. Yaresko, T. K. Kim, M. Hoesch, T. Wolf, and N. D. Zhigadlo, Nat. Phys., \textbf{12}, 311 (2016).
\bibitem{Watson} M. D. Watson, A. A. Haghighirad, H. Takita, W. Mansuer, H. Iwasawa,
E. F. Schwier, A. Ino, and M. Hoesch, J. Phys. Soc. Jpn., \textbf{86}, 053703 (2017).
\bibitem{Chubukov4} The same approximation has been done on a similar itinerant model by other authors in the literature (see, e.g., Ref. \cite{Chubukov}) and was shown not to affect the main conclusions of the RG study.
\bibitem{Peskin} M. E. Peskin and D. V. Schroeder, \emph{An Introduction to Quantum Field Theory} (Addison-Wesley, Reading, 1995).
\bibitem{Freire3} H. Freire, V. S. de Carvalho, and C. Pépin, Phys. Rev. B \textbf{92}, 045132 (2015).
\bibitem{Freire4} V. S. de Carvalho and H. Freire, Annals of Physics \textbf{348}, 32 (2014); V. S. de Carvalho and H. Freire, Nucl. Phys. B \textbf{875}, 738 (2013).
\bibitem{Bychkov} Y. A. Bychkov, L. P. Gor'kov, and I. E. Dzyaloshinskii, Sov. Phys. JETP \textbf{23}, 489 (1966).
\bibitem{Chubukov5} A. V. Chubukov, R. M. Fernandes, and J. Schmalian, Phys. Rev. B \textbf{91}, 201105
(2015).
\bibitem{Mazin} I. I. Mazin, D. J. Singh, M. D. Johannes, and M. H. Du, Phys. Rev. Lett.
\textbf{101}, 057003 (2008).
\bibitem{Chubukov6} A. V. Chubukov, D. V. Efremov, and I. Eremin, Phys. Rev. B \textbf{78}, 134512 (2008).
\bibitem{Kontani} Y. Yamakawa, S. Onari, and H. Kontani,
Phys. Rev. X \textbf{6}, 021032 (2016).
\bibitem{Fanfarillo} L. Fanfarillo, J. Mansart, P. Toulemonde, H. Cercellier,  P. Le Fèvre, F. Bertran, B. Valenzuela, L. Benfatto, and V. Brouet, Phys. Rev. B \textbf{94}, 155138 (2016)
\bibitem{Si2} R. Yu and Q. Si, Phys. Rev. Lett. \textbf{115}, 116401 (2015).
\bibitem{Nevidomskyy} Z. Wang, W.-J. Hu, and A. H. Nevidomskyy, Phys. Rev. Lett. \textbf{116}, 247203 (2016)
\bibitem{DHLee} F. Wang, S. A. Kivelson, and D.-H. Lee, Nat. Phys. \textbf{11}, 959 (2015).
\bibitem{Si3} H. Hu, R. Yu, E. M. Nica, J.-X. Zhu, and Q. Si, arXiv:1805.05915 (2018).
\end{thebibliography}
\end{document}